\documentclass[12pt]{article}
\usepackage{amssym}
\usepackage{graphics}

\def\CS{|z\rangle_q}
\def\ap{a^{\dagger}}
\def\C{\mbox{$\Bbb C$}}
\def\tW{\tilde{W}}
\def\case#1#2{{\textstyle{#1\over #2}}}
\def\bp{b^{\dagger}}

\title{
New $q$-deformed coherent states with an explicitly known resolution of unity}
\author{C. Quesne \thanks{Directeur de recherches FNRS; E-mail address:
cquesne@ulb.ac.be}\\ 
{\small Physique Nucl\'eaire Th\'eorique et Physique
Math\'ematique,  Universit\'e Libre de Bruxelles,} \\ 
{\small Campus de la Plaine CP229, Boulevard~du Triomphe, B-1050 Brussels,
Belgium}}
\date{ }
\begin{document}
\baselineskip=20pt plus 1pt minus 1pt
\maketitle

\begin{abstract} 
We construct a new family of $q$-deformed coherent states $\CS$, where $0 < q < 1$.
These states are normalizable on the whole complex plane and continuous in their label
$z$. They allow the resolution of unity in the form of an ordinary integral with a positive
weight function obtained through the analytic solution of the associated Stieltjes
power-moment problem and expressed in terms of one of the two Jacksons's
$q$-exponentials. They also permit exact evaluation of matrix elements of
physically-relevant operators. We use this to show that the photon number statistics for
the states is sub-Poissonian and that they exhibit quadrature squeezing as well as an
enhanced signal-to-quantum noise ratio over the conventional coherent state value.
Finally, we establish that they are the eigenstates of some deformed boson annihilation
operator and study some of their characteristics in deformed quantum optics. 
\end{abstract}
%
%
\newpage
\section{Introduction}
\label{sec:intro}

The conventional coherent states (CS)~\cite{glauber, klauder63, sudarshan}, dating back
to the birth of quantum mechanics~\cite{schroedinger}, are specific superpositions
$|z\rangle$, parametrized by a single complex number $z$, of the eigenstates
$|n\rangle$ of the harmonic oscillator number operator $N = \ap a$. They have
properties similar to those of the classical radiation field and may be constructed in three
equivalent ways: (i) by defining them as eigenstates of the annihilation operator $a$, (ii)
by applying a unitary displacement operator on the vacuum state $|0\rangle$ (such that
$a |0\rangle = 0$), and (iii) by considering them as quantum states with a minimum
uncertainty relationship.\par
%
%
In recent years, other classes of states often termed `nonclassical' have attracted a lot
of attention in quantum optics, as well as in other areas ranging from solid state physics
to cosmology. Such states exhibit some purely quantum-mechanical properties, such as
squeezing, higher-order squeezing, antibunching, sub-Poissonian  statistics, etc (for a
review see~\cite{dodonov} and references quoted therein). Among them, one finds the
so-called generalized coherent states (GCS), arising in connection with algebras other
than the oscillator one~\cite{klauder85}. They may be defined by extending one of the
three equivalent definitions of the conventional CS, which become inequivalent for more
complicated algebras. One may therefore distinguish between
annihilation-operator~\cite{barut}, displacement-operator~\cite{perelomov}, and
minimum-uncertainty or intelligent GCS~\cite{aragone}.\par
%
%
{}From a mathematical viewpoint, it has been noted~\cite{klauder63} that to be
acceptable GCS should satisfy a minimum set of conditions: (i) normalizability (as any
vector of a Hilbert space), (ii) continuity in the label $z$, and (iii) existence of a
resolution of unity with a positive-definite weight function (implying that the states form
an (over)complete set). If the first two conditions are rather easy to fulfil, the third one,
on the contrary, imposes some severe restrictions on possible GCS. Determining the
existence of a unity resolution relation is indeed a difficult task, which has not been
solved for many of the putative GCS known in the literature. Recent progress in this
domain has been achieved by using some properties of the inverse Mellin
transform~\cite{sixdeniers,cq}.\par
%
%
A class of GCS, which has evoked a lot of interest, is connected with deformations of the
canonical commutation relations or, equivalently, with deformed oscillator algebras. Since
the introduction of $q$-deformed CS~\cite{arik, jannussis, biedenharn} associated with
the so-called {\em maths}-type~\cite{arik, kuryshkin} and {\em
physics}-type~\cite{biedenharn, macfarlane} $q$-deformed bosons, various
generalizations have been considered (for a review see~\cite{dodonov}), including
$q$-deformed CS associated with two-parameter deformations~\cite{chakrabarti}. This
has culminated in the definition of `nonlinear' CS~\cite{matos}, for which the deformation
is governed by some function of $N$.\par
%
%
{}Few $q$-deformed CS have been endowed with a unity resolution relation. For those
connected with maths-type and physics-type $q$-deformed bosons, the latter has been
written as a $q$-integral~\cite{jackson} with a weight function expressed in terms of some
$q$-exponential~\cite{arik, gray}. An alternative formulation has been proposed in terms
of an ordinary integral, but the corresponding weight function is then only known through
the inverse Fourier transform of some given function~\cite{kar}. Another type of
$q$-deformed CS has also been shown to admit a unity resolution relation in the form of
an ordinary integral with a weight function expressed as a Laplace
transform of some given function~\cite{penson}.\par
%
%
The purpose of the present paper is to prove that in contrast with known
$q$-deformed CS, it is possible to build some new ones admitting a unity resolution
relation expressed in terms of an ordinary integral with an explicitly known positive weight
function. Moreover such GCS will be shown to possess some interesting nonclassical
properties.\par
%
%
In section~\ref{sec:new}, new $q$-deformed physical states are introduced and shown to
qualify as GCS in the sense described above. Their geometrical and physical properties in
quantum optics are studied in section~\ref{sec:prop}. In section~\ref{sec:deformed},
their description in terms of $q$-deformed bosons is considered. Finally,
section~\ref{sec:conclusion} contains the conclusion.\par
%
%
\section{\boldmath New $q$-deformed coherent states in terms of boson operators}
\label{sec:new}

Let us consider a new family of harmonic oscillator physical states
\begin{equation}
  \CS = {\cal N}_q^{-1/2}(|z|^2) \sum_{n=0}^{\infty} \frac{z^n}{\sqrt{[n]_q!}} \,
  |n\rangle  \label{eq:CS}
\end{equation}
labelled by $q \in (0,1)$ and $z \in \C$. Here $|n\rangle = (n!)^{-1/2} (\ap)^n
|0\rangle$ is an $n$-boson state and
\begin{eqnarray}
  [n]_q! & \equiv & [n]_q [n-1]_q \ldots [1]_q \qquad {\rm if\ } n = 1, 2, \ldots
        \nonumber \\
  & \equiv & 1 \qquad {\rm if\ } n = 0  \label{eq:q-fac}
\end{eqnarray}
where
\begin{equation}
  [n]_q \equiv \frac{1-q^{-n}}{q-1} = q^{-n} \{n\}_q  \label{eq:q-number}
\end{equation}
differs from the corresponding quantity $\{n\}_q \equiv (1-q^n)/(1-q)$ for maths-type
$q$-deformed bosons~\cite{arik, jannussis, kuryshkin} by the $n$- and $q$-dependent
factor $q^{-n}$. Note that the $q$-factorial (\ref{eq:q-fac}) can also be written in terms
of the $q$-gamma function $\Gamma_q(x)$~\cite{gasper} as
\begin{equation}
  [n]_q! = q^{-n(n+1)/2} \Gamma_q(n+1).  \label{eq:q-gamma}
\end{equation}
For $q \to 1^-$, $[n]_q$, $[n]_q!$, and $\Gamma_q(n)$ go to $n$, $n!$, and
$\Gamma(n)$, respectively.\par
%
%
Let us now show that the states $|z\rangle_q$ are GCS in the sense described in
section~\ref{sec:intro}.\par
%
%
\subsection{Normalizability}
\label{subsec:norm}

The normalization condition of (\ref{eq:CS}), ${}_q\langle z|z \rangle_q = 1$, leads to
\begin{equation}
  {\cal N}_q(x) = \sum_{n=0}^{\infty} \frac{x^n}{[n]_q!} = \sum_{n=0}^{\infty}
  \frac{q^{n(n-1)/2}}{\{n\}_q!} (qx)^n = E_q[(1-q)qx]  \label{eq:norm} 
\end{equation}
where $x \equiv |z|^2$, $\{n\}_q!$ is defined in a way similar to $[n]_q!$, and $E_q(z)
\equiv \prod_{k=0}^{\infty} (1 + q^k z)$ is one of the two $q$-exponentials introduced
by Jackson (such that $\lim_{q\to 1^-} E_q[(1-q)z] = \exp(z)$)~\cite{jackson}.
Since $E_q(z)$ converges absolutely for $0 < |q| < 1$ and any $z$~\cite{gasper}, ${\cal
N}_q(x)$ is an entire function. Moreover it is positive on the positive real axis. The states
(\ref{eq:CS}) are therefore normalizable on the whole complex plane. For $q \to 1^-$,
they reduce to the conventional CS~\cite{glauber, klauder63, sudarshan}
\begin{equation}
  |z\rangle = {\cal N}^{-1/2}(x) \sum_{n=0}^{\infty} \frac{z^n}{\sqrt{n!}}\, |n\rangle
  \qquad {\cal N}(x) = e^x.  \label{eq:conv-CS}
\end{equation}
\par
%
%
A convenient way to visualize how $\CS \stackrel{q\to 1}{\longrightarrow} [z\rangle$ is
to display the function
\begin{equation}
  d_q(n) = \frac{[n]_q!}{n!}  \label{eq:d}
\end{equation}
for different values of the parameter $q$. This is done in figure~1. We note that
$d_q(n)$ is always greater than one. For a given $q$ it is an increasing function of $n$,
while for a given $n$ its departure from one increases for decreasing $q$ values. It
follows that for such $q$ values the convergence of the series defining ${\cal N}_q(x)$ in
(\ref{eq:norm}) becomes increasingly fast as compared with that defining the
exponential. This is confirmed in figure~2, where ${\cal N}_q(x)$ is displayed for several
$q$ values and compared with ${\cal N}(x)$. Observe that for $x \ll 1$, ${\cal N}_q(x)$
behaves as
\begin{equation}
  {\cal N}_q(x) \simeq 1 + qx + O(x^2).
\end{equation}
\par
%
%
\subsection{\boldmath Continuity in $z$}
\label{subsec:continuity}

The states $\CS$, defined in (\ref{eq:CS}), are continuous in $z$ if
\begin{equation}
  [z - z'| \to 0 \Rightarrow \Bigl| \CS - |z'\rangle_q \Bigr|^2 \to 0.  \label{eq:continuous}
\end{equation}
Since
\begin{equation}
  \Bigl| \CS - |z'\rangle_q \Bigr|^2 = 2 (1 - {\rm Re}\, {}_q \langle z|z' \rangle_q)
\end{equation}
where
\begin{equation}
  {}_q \langle z|z' \rangle_q = \left[{\cal N}_q(|z|^2) {\cal N}_q(|z'|^2)\right]^{-1/2}
  E_q[(1-q)q z^{\prime*}z]   
\end{equation}
is a continuous function, it follows that condition (\ref{eq:continuous}) is satisfied.\par
%
%
\subsection{Resolution of unity}
\label{sec:unity}

The states $\CS$ give rise to a resolution of unity with weight function $W_q(x)$ if
\begin{equation}
  \int\int_{\C} d^2z\, \CS\, W_q(|z|^2)\, {}_q\langle z| = \sum_{n=0}^{\infty} |n\rangle
  \langle n| = I.
\end{equation}
On using expansions of $\CS$ and ${}_q \langle z|$ in terms of $n$-boson kets and bras,
such a condition is transformed into
\begin{equation}
  \sum_{n=0}^{\infty} \left(\frac{\pi}{[n]_q!} \int_0^{\infty} dx\, \frac{x^n W_q(x)}
  {{\cal N}_q(x)}\right) |n\rangle \langle n| = I.
\end{equation}
It can be satisfied provided the function
\begin{equation}
  \tW_q(x) \equiv \pi \frac{W_q(x)}{{\cal N}_q(x)}
\end{equation}
is a solution of the following Stieltjes power-moment problem:
\begin{equation}
  \int_0^{\infty} dx\, x^n \tW_q(x) = [n]_q! \qquad n=0, 1, 2, \ldots.  \label{eq:power}
\end{equation}
\par
%
%
{}From the relation (\ref{eq:q-gamma}) between the $q$-factorial and the $q$-gamma
function and from a $q$-analogue of the Euler gamma integral~\cite{ataki}, we directly
obtain a solution of (\ref{eq:power}) in the form
\begin{equation}
  \tW_q(x) = \frac{1-q}{\ln q^{-1}} \{E_q[(1-q)x]\}^{-1}  \label{eq:tW}
\end{equation}
from which the relation
\begin{equation}
  W_q(x) = \frac{1-q}{\pi \ln q^{-1}} \frac{E_q[(1-q)qx]}{E_q[(1-q)x]}  \label{eq:W} 
\end{equation}
follows.\par
%
%
It is easy to check that in the limit $q \to 1^-$, $\tW_q(x) \to \tW(x) = e^{-x}$ and
$W_q(x) \to W(x) = \pi^{-1}$, thus giving back the corresponding results for the
conventional CS (\ref{eq:conv-CS})~\cite{glauber, klauder63, sudarshan}. For $x \ll 1$,
we get
\begin{equation}
  \tW_q(x) \simeq \frac{1-q}{\ln q^{-1}} [1 - x + O(x^2)] \qquad W_q(x) \simeq
  \frac{1-q}{\pi \ln q^{-1}} [1 - (1-q)x + O(x^2)].  
\end{equation}
In figures 3 and 4, the weight functions $\tW_q(x)$ and $W_q(x)$ are displayed for
several $q$ values and compared with $\tW(x)$ and $W(x)$, respectively.\par
%
%
We therefore conclude that the new physical states $\CS$, defined in (\ref{eq:CS}),
qualify as GCS in the sense described in section~\ref{sec:intro}.\par
%
%
It is worth emphasizing however that in contradistinction to the case of the conventional
CS, the solution $\tW_q(x)$ of the power-moment problem (\ref{eq:power}) may not be
unique as the so-called Carleman criterion~\cite{akhiezer} indicates. According to the
latter, if a solution exists and $S \equiv \sum_{n=1}^{\infty} a_n$, where $a_n \equiv
([n]_q!)^{-1/(2n)}$, diverges (resp.\ converges), then the solution is unique (resp.\
nonunique solutions may exist). The logarithmic test~\cite{prudnikov}, $\lim_{n\to\infty}
(\ln a_n/\ln n) = \lim_{n\to\infty} (4\ln n)^{-1} (n+1) \ln q < -1$, actually shows the
convergence of $S$, hence the possible existence of other solutions.\par
%
%
\section{Geometrical and physical properties in quantum optics}
\label{sec:prop}
\setcounter{equation}{0}

In the present section, we shall proceed to study some geometrical and physical
properties of our new $q$-deformed CS $\CS$. For such a purpose, we shall need to
evaluate the expectation values of some Hermitian monomials in the boson creation and
annihilation operators $\ap$, $a$. These are expressible through the derivatives of ${\cal
N}_q(x)$ as~\cite{sixdeniers, penson}
\begin{equation}
  {}_q \langle (\ap)^r a^r\rangle_q = \frac{x^r}{{\cal N}_q(x)} \frac{d^r 
  {\cal N}_q(x)}{dx^r} \qquad r=0, 1, 2, \dots  \label{eq:average}
\end{equation}
where we use the notation
\begin{equation}
  {}_q \langle O \rangle_q \equiv {}_q \langle z|O\CS.
\end{equation}
More generally, for non-Hermitian monomials, equation~(\ref{eq:average}) can be
extended to
\begin{equation}
  {}_q \langle (\ap)^p a^r\rangle_q = (z^*)^p z^r S^{(p,r)}_q(x)
\end{equation}
where
\begin{equation}
  S^{(p,r)}_q(x) = \frac{1}{{\cal N}_q(x)} \sum_{n=0}^{\infty} \left(\frac{(n+p)! (n+r)!}
  {[n+p]_q! [n+r]_q!}\right)^{1/2} \frac{x^n}{n!} \qquad r, p = 0, 1, 2, \ldots.
\end{equation}
\par
%
%
\subsection{Metric factor}
\label{subsec:metric}

The map from $z$ to $\CS$ is a map from \C\ into a continuous subset of unit vectors
in Hilbert space. It generates in the latter a two-dimensional surface, whose non-flat,
circularly symmetric geometry~\cite{sixdeniers} is described in polar coodinates $r$,
$\theta$ (i.e., $z = r e^{{\rm i}\theta}$) by the line element $d\sigma_q^2 =
\omega_q(r^2) (dr^2 + r^2 d\theta^2)$, where the metric factor is 
\begin{equation}
  \omega_q(x) = \frac{d}{dx} \langle N\rangle_q = \left(\frac{x {\cal N}'_q(x)}
  {{\cal N}_q(x)}\right)' = \frac{{\cal N}'_q(x)}{{\cal N}_q(x)} + x \left[\frac{{\cal
  N}''_q(x)}{{\cal N}_q(x)} - \left(\frac{{\cal N}'_q(x)}{{\cal N}_q(x)}\right)^2\right]. 
  \label{eq:metric}
\end{equation}
Here the primes denote the order of differentiations with respect to the variable $x$. One
can interpret $d\sigma_q^2$ as the geometry of the associated classical phase
space.\par
%
%
{}For $x \ll 1$, we obtain
\begin{equation}
  \omega_q(x) \simeq q \left[1 - \frac{2q(1-q)}{1+q} x + O(x^2)\right].
\end{equation}
Figure 5, which shows $\omega_q(x)$ for several $q$ values, confirms that over an
extended $x$ range, $\omega_q(x) < \omega(x) = 1$ corresponding to the flat
geometry of conventional CS.\par
%
%
\subsection{Photon number distribution}
\label{subsec:mandel}

The probability of finding $n$ bosons in the $q$-deformed CS $\CS$ is equal to
\begin{equation}
  p_q(n,x) = \frac{x^n}{{\cal N}_q(x) [n]_q!}.
\end{equation}
In the limit $q \to 1^-$, it reduces to a Poisson distribution for the conventional CS.\par
%
%
Since for the latter the variance of the number operator $N$ is equal to its average,
deviations from Poisson statistics can be measured with the Mandel
parameter~\cite{mandel}
\begin{equation}
  Q_q(x) = \frac{(\Delta N)_q^2 - \langle N\rangle_q}{\langle N\rangle_q} \qquad
  (\Delta N)_q^2 \equiv \langle N^2\rangle_q - \langle N\rangle_q^2
\end{equation}
which vanishes for the Poisson distribution, is positive for a super-Poissonian distribution
(bunching effect), and is negative for a sub-Poissonian distribution (antibunching effect).
From (\ref{eq:average}), it follows that $Q_q(x)$ can be written as
\begin{equation}
  Q_q(x) = x \left(\frac{{\cal N}''_q(x)}{{\cal N}'_q(x)} - \frac{{\cal N}'_q(x)}{{\cal
  N}_q(x)}\right).  \label{eq:mandel} 
\end{equation}
\par
%
%
{}For $x \ll 1$, we obtain
\begin{equation}
  Q_q(x) \simeq - \frac{q(1-q)}{1+q} x + O(x^2).
\end{equation}
This result hints at a sub-Poissonian distribution. Numerical study confirms that this is
indeed true, as displayed in figure 6.\par
%
%
\subsection{Squeezing properties}
\label{subsec:squeezing}

Let us consider the Hermitian quadrature operators
\begin{equation}
  X = \frac{1}{\sqrt{2}} (a + \ap) \qquad P = \frac{1}{{\rm i}\sqrt{2}} (a - \ap). 
\end{equation}
Their variances $(\Delta X)^2$ and $(\Delta P)^2$ in any state $|\psi\rangle$ satisfy
the conventional uncertainty relation
\begin{equation}
  (\Delta X)^2 (\Delta P)^2 \ge \case{1}{4}
\end{equation}
the lower bound being attained by the vacuum state, for which both variances are equal
to $\frac{1}{2}$. A state $|\psi\rangle$ is said to be squeezed for the quadrature $X$
(resp.\ $P$) if $(\Delta X)^2 < \frac{1}{2}$ (resp.\ $(\Delta P)^2 < \frac{1}{2}$).\par
%
%
{}For the $q$-deformed CS $\CS$, it is straightforward to show that the variances of
$X$ and $P$ are given by
\begin{eqnarray}
  (\Delta X)_q^2 & = & 2({\rm Re}\, z)^2 \left\{S^{(2,0)}_q(x) - \left[S^{(1,0)}_q(x)
         \right]^2\right\} + x \left[S^{(1,1)}_q(x) - S^{(2,0)}_q(x)\right] + \frac{1}{2}
         \label{eq:variance}\\
  (\Delta P)_q^2 & = & 2({\rm Im}\, z)^2 \left\{S^{(2,0)}_q(x) - \left[S^{(1,0)}_q(x)
         \right]^2\right\} + x \left[S^{(1,1)}_q(x) - S^{(2,0)}_q(x)\right] + \frac{1}{2}.
\end{eqnarray}
Since they are related to each other by the transformation ${\rm Re}\, z \leftrightarrow
{\rm Im}\, z$, it is enough to study the former.\par
%
%
{}For small $x$ values, the coefficient of $2 ({\rm Re}\, z)^2$ in (\ref{eq:variance})
behaves as $q \left(\sqrt{\frac{2q}{1+q}} - 1\right)$, which is negative for $0 < q < 1$.
It can be checked numerically that this property holds true over a wide range of $x$
values. Hence the maximum squeezing in $X$ can be achieved when $z$ is real.\par
%
%
{}For $0 < z = \sqrt{x} \ll 1$, the ratio
\begin{equation}
  R_q(x) = 2 (\Delta X)_q^2  \label{eq:R}
\end{equation}
of the variance $(\Delta X)_q^2$ in $\CS$ to the variance $\frac{1}{2}$ in the vacuum
state behaves as
\begin{equation}
  R_q(x) \simeq 1 + 2q \left(\sqrt{\frac{2q}{1+q}} - 1\right) x + O(x^2)
\end{equation}
which shows that squeezing is present for small $x$ values. This is confirmed by
figure~7, where $R_q(x)$ is plotted against $x$ for $z = \sqrt{x}$ and several $q$
values. We conclude that there is a substantial squeezing increasing with $x$ and with
the deformation.\par
%
%
\subsection{Signal-to-quantum noise ratio}
\label{subsec:signal}

{}Finally, let us calculate the signal-to-quantum noise ratio $\sigma_q$, which is defined
by
\begin{equation}
  \sigma_q = \frac{\langle X\rangle_q^2}{(\Delta X)_q^2}  \label{eq:signal}
\end{equation}
where $(\Delta X)_q^2$ is given in (\ref{eq:variance}) and
\begin{equation}
  \langle X\rangle_q^2 = 2 ({\rm Re}\, z)^2 \left[S^{(1,0)}_q(x)\right]^2.  
\end{equation}
It follows that for a given $x$ value, such a ratio is maximum for real $z$. For $0 < z =
\sqrt{x} \ll 1$, it behaves as
\begin{equation}
  \sigma_q(x) \simeq 4qx [1 + O(x^2)].  
\end{equation}
\par
%
%
{}For the conventional CS, the ratio $\sigma$ ($= \sigma_1$) attains the value $4N_s$,
where $N_s = \langle N\rangle$ is the number of photons in the signal. In the case of the
$q$-deformed CS $\CS$, we may compare $\sigma_q(x)$ to $4 \langle N \rangle_q$.
For $x \ll 1$, the latter behaves as
\begin{equation}
  4 \langle N \rangle_q \simeq 4qx \left[1 - \frac{q(1-q)}{1+q} x + O(x^2)\right].  
\end{equation}
On comparing with (\ref{eq:signal}), it is clear that there is an improvement on the
conventional CS value.\par
%
%
We may also compare $\sigma_q(x)$ to the upper bound $4N_s(N_s+1) = 4 \langle
N\rangle_q (\langle N\rangle_q + 1)$~\cite{yuen}, which is attained by a conventional
squeezed state~\cite{hollenhorst}. For $x \ll 1$, we obtain
\begin{equation}
  4 \langle N\rangle_q (\langle N\rangle_q + 1) \simeq 4qx \left[1 + \frac{2q^2}{1+q} x 
  + O(x^2)\right]
\end{equation}
thus confirming the prediction.\par
%
%
As displayed in figure~8, numerical calculations show the validity of the above conclusions
over an extended range of $x$ values.\par
%
%
\section{\boldmath Description in terms of $q$-deformed bosons}
\label{sec:deformed}
\setcounter{equation}{0}

\subsection{\boldmath The new $q$-deformed coherent states in terms of
$q$-deformed bosons}
\label{subsec:q-bosons}

The new $q$-deformed CS, introduced in section~\ref{sec:new} in terms of conventional
boson operators $\ap$, $a$, can alternatively be described in terms of some
$q$-deformed boson operators $\bp$, $b$, defined in terms of the latter by
\begin{equation}
  \bp = \sqrt{\frac{[N]_q}{N}}\, \ap \qquad b = a \sqrt{\frac{[N]_q}{N}} 
  \label{eq:q-boson}
\end{equation}
where $[N]_q$ is given, as in (\ref{eq:q-number}), by
\begin{equation}
  [N]_q = \frac{1-q^{-N}}{q-1}.
\end{equation}
Contrary to the boson operators for which $\ap a = N$, $a\ap =N+1$, the $q$-deformed
ones are such that
\begin{equation}
  \bp b = [N]_q \qquad b \bp = [N+1]_q.
\end{equation}
\par
%
%
{}From these relations, it follows that they satisfy the commutation relation
\begin{equation}
  [b, \bp] = [N+1]_q - [N]_q = q^{-N-1}  \label{eq:q-com}
\end{equation}
together with
\begin{equation}
  [N, \bp] = \bp \qquad [N, b] = -b  \label{eq:N-b}
\end{equation}
resulting from (\ref{eq:q-boson}). Observe that the $N$-dependent commutation relation
(\ref{eq:q-com}) may be replaced by an $N$-independent one provided we substitute a
`quommutator' for the ordinary commutator:
\begin{equation}
  [b, \bp]_{q^{-1}} \equiv b\bp - q^{-1} \bp b = q^{-1}.
\end{equation}
\par
%
%
The operators $\bp$, $b$ act on the same Fock space $\{|n\rangle \mid n=0, 1, 2,
\ldots\}$ as $\ap$, $a$. From (\ref{eq:q-boson}), we obtain
\begin{equation}
  b |n\rangle = \sqrt{[n]_q}\, |n-1\rangle \qquad \bp |n\rangle = \sqrt{[n+1]_q}\,
  |n+1\rangle  \label{eq:b-action}
 \end{equation}
with $b |0\rangle = 0$. The $n$-boson states $|n\rangle$ can therefore be rewritten as
$n$-deformed-boson states through the relation
\begin{equation}
  |n\rangle = \frac{1}{\sqrt{[n]_q!}}\, (\bp)^n |0\rangle.  \label{eq:n-boson}
\end{equation}
\par
%
%
On inserting equation (\ref{eq:n-boson}) in (\ref{eq:CS}) and using the definition of the
$q$-exponential given in (\ref{eq:norm}), we can express our new $q$-deformed CS in
terms of the $q$-deformed boson operators $\bp$, $b$ as
\begin{equation}
  \CS = {\cal N}_q^{-1/2}(x) E_q[(1-q)qz\bp] |0\rangle.  \label{eq:CS-bis}
\end{equation}
From (\ref{eq:CS}) and (\ref{eq:b-action}), it also follows that
\begin{equation}
  b \CS = z \CS  \label{eq:annihilation}
\end{equation}
showing that the $q$-deformed CS $\CS$ are annihilation-operator CS for the
$q$-deformed oscillator algebra generated by $N$, $\bp$, $b$, and defined by equations
(\ref{eq:q-com}) and (\ref{eq:N-b}). In the limit $q \to 1^-$, equations (\ref{eq:CS-bis})
and (\ref{eq:annihilation}) reduce to known results for the conventional CS, namely
$|z\rangle = {\cal N}^{-1/2}(x) \exp(z\ap) |0\rangle$ and $a |z\rangle = z
|z\rangle$.\par
%
%
Observe that as announced in section~\ref{sec:intro}, the states (\ref{eq:CS}) or
(\ref{eq:CS-bis}) cannot be obtained by applying a unitary displacement operator on the
vacuum if the base field remains the complex plane. It might however prove possible to
construct such a displacement operator by using a non-commutative base field, as was
done in~\cite{mcdermott} for the maths-type $q$-deformed CS.\par
%
%
\subsection{Physical properties in deformed quantum optics}
\label{subsec:def-QO}

The $q$-deformed boson operators $\bp$, $b$ may be interpreted as describing
`dressed' photons, which may be invoked in phenomenological models explaining some
non-intuitive observable phenomena~\cite{katriel}. The physical applications considered in
section~\ref{sec:prop} for `real' photons may therefore be re-examined for those
deformed photons.\par
%
%
In such a context, it has been shown~\cite{solomon} that the properties of
$q$-deformed CS, defined as in (\ref{eq:annihilation}), are governed by a characteristic
functional
\begin{equation}
  \rho_q(n) = \frac{[n+1]_q/[n]_q}{(n+1)/n}
\end{equation}
which takes the value one for conventional photons. In the present case, it can easily be
proved that $\rho_q(n) > 1$ for $0 < q < 1$.\par
%
%
{}From this, we deduce the following consequences:

\noindent
(i) The Mandel parameter $Q_q(x)$ is negative, which corresponds to a sub-Poissonian
distribution. This agrees with the results of section~\ref{subsec:mandel}.

\noindent
(ii) The variances of the deformed quadrature operators
\begin{equation}
  X_b = \frac{1}{\sqrt{2}} (b + \bp) \qquad P_b = \frac{1}{{\rm i}\sqrt{2}} (b - \bp) 
\end{equation}
in any state satisfy the uncertainty relation
\begin{equation}
  (\Delta X_b)^2 (\Delta P_b)^2 \ge \case{1}{4} |\langle[X_b, P_b]\rangle|^2.
\end{equation}
For the $q$-deformed CS $\CS$, such a relation becomes an equality since
\begin{eqnarray}
  (\Delta X_b)_q^2 & = & (\Delta P_b)_q^2 = \case{1}{2} |\langle[X_b, P_b]\rangle_q| =
       \case{1}{2} (\langle [N+1]_q \rangle_q - \langle [N]_q \rangle_q) \nonumber \\
  & = & \case{1}{2} \left\{\frac{1}{{\cal N}_q(x)} \sum_{n=0}^{\infty} [n+1]_q
\frac{x^n}{[n]_q!} - x\right\}. \label{eq:b-variance}  
\end{eqnarray}
The states $\CS$ are therefore intelligent states for the deformed operators $X_b$,
$P_b$. However they are not minimum-uncertainty states as the vacuum uncertainty
product provides a global lower bound. Moreover no squeezing occurs in $\CS$ since the
ratio 
\begin{equation}
  R_{bq}(x) = 2q (\Delta X_b)_q^2  \label{eq:Rb} 
\end{equation}
of $(\Delta X_b)_q^2$ to the variance $1/(2q)$ of $X_b$ in the vacuum state is
always greater than one. This is illustrated in figure~9, where $R_{bq}(x)$ is plotted
against
$x$ for several $q$ values.

\noindent
(iii) The signal-to-quantum noise ratio for deformed photons
\begin{equation}
  \sigma_{bq} = \frac{\langle X_b\rangle_q^2}{(\Delta X_b)_q^2}  \label{eq:signal-b}
\end{equation}
is not enhanced over the conventional photon case. Here $(\Delta X_b)_q^2$ is given
by (\ref{eq:b-variance}) and $\langle X_b\rangle_q = \sqrt{2}\, {\rm Re}\, z$, hence, for
a given $x$ value, $\sigma_{bq}$ is maximum for real $z$. On figure~8, it can be seen
that for such values, $\sigma_{bq}$ is indeed lower than $4 \langle N\rangle_q$.\par
%
%
\section{Conclusion}
\label{sec:conclusion}

In the present paper, we have defined new $q$-deformed CS by slightly modifying the
maths-type $q$-deformed CS~\cite{arik, jannussis}. Such a change has allowed us to give
the explicit weight function that satisfies the associated Stieltjes power-moment problem
and thus the unity resolution property in the form of an ordinary integral.\par
%
%
The simple form of our CS has also enabled us to provide analytical formulae for
calculating matrix elements of physically-relevant operators. With their use, we have
investigated some characteristics of our CS relevant to quantum optics. We have
established that they exhibit strong non-classical properties, such as antibunching,
quadrature squeezing, and enhancement of the signal-to-quantum noise ratio.\par
%
%
{}Finally, we have shown that this family of states can be expressed as eigenstates of
some deformed boson annihilation operator and we have studied some phenomena in
deformed quantum optics.\par
%
%
\section*{Acknowlegments}

The author thanks K.\ Penson for an interesting discussion and B.\ Bagchi for informing
her about~\cite{dodonov}.\par
%
%
 
%
%
\newpage
\section*{Figure captions}

{\bf Figure 1.} The ratio $d_q(n)$ of equation~(\ref{eq:d}) versus $n$ for $q=0.98$
(rhombs), $q=0.96$ (stars), and $q=0.94$ (squares).

\noindent
{\bf Figure 2.} The normalization function ${\cal N}_q(x)$ of equation~(\ref{eq:norm})
versus $x$ for $q=1$ (solid line), $q=0.9$ (dashed line), $q=0.8$ (dotted line), and
$q=0.7$ (dot-dashed line).

\noindent
{\bf Figure 3.} The weight function $\tW_q(x)$ of equation~(\ref{eq:tW}) versus $x$ for
$q=1$ (solid line), $q=0.9$ (dashed line), $q=0.8$ (dotted line), and $q=0.7$ (dot-dashed line).

\noindent
{\bf Figure 4.} The weight function $W_q(x)$ of equation~(\ref{eq:W}) versus $x$ for
$q=1$ (solid line), $q=0.9$ (dashed line), $q=0.8$ (dotted line), and $q=0.7$
(dot-dashed line).

\noindent
{\bf Figure 5.} The metric factor $\omega_q(x)$ of equation~(\ref{eq:metric}) versus
$x$ for $q=1$ (solid line), $q=0.9$ (dashed line), $q=0.8$ (dotted line), and $q=0.7$
(dot-dashed line).

\noindent
{\bf Figure 6.} The Mandel parameter $Q_q(x)$ of equation~(\ref{eq:mandel}) versus $x$
for $q=1$ (solid line), $q=0.9$ (dashed line), $q=0.8$ (dotted line), and $q=0.7$
(dot-dashed line).

\noindent
{\bf Figure 7.} The variance ratio $R_q(x)$ of equation~(\ref{eq:R}) versus $x$ for
 real $z = \sqrt{x}$ and $q=1$ (solid line), $q=0.9$ (dashed line), $q=0.8$ (dotted
line), or $q=0.7$ (dot-dashed line).

\noindent
{\bf Figure 8.} Signal-to-quantum noise ratio plots versus $x$ for $q=0.7$: $\sigma_q$
of equation~(\ref{eq:signal}) for real $z = \sqrt{x}$ (solid line), $4\langle N\rangle_q$
(dashed line), $4\langle N\rangle_q (\langle N\rangle_q + 1)$ (dotted line), and
$\sigma_{bq}$ of equation~(\ref{eq:signal-b}) (dot-dashed line).

\noindent
{\bf Figure 9.} The variance ratio $R_{bq}(x)$ of equation~(\ref{eq:Rb}) versus $x$ for
$q=1$ (solid line), $q=0.9$ (dashed line), $q=0.8$ (dotted line), and $q=0.7$
(dot-dashed line).
%
%
\newpage
\begin{picture}(160,100)
\put(35,0){\mbox{\scalebox{1.0}{\includegraphics{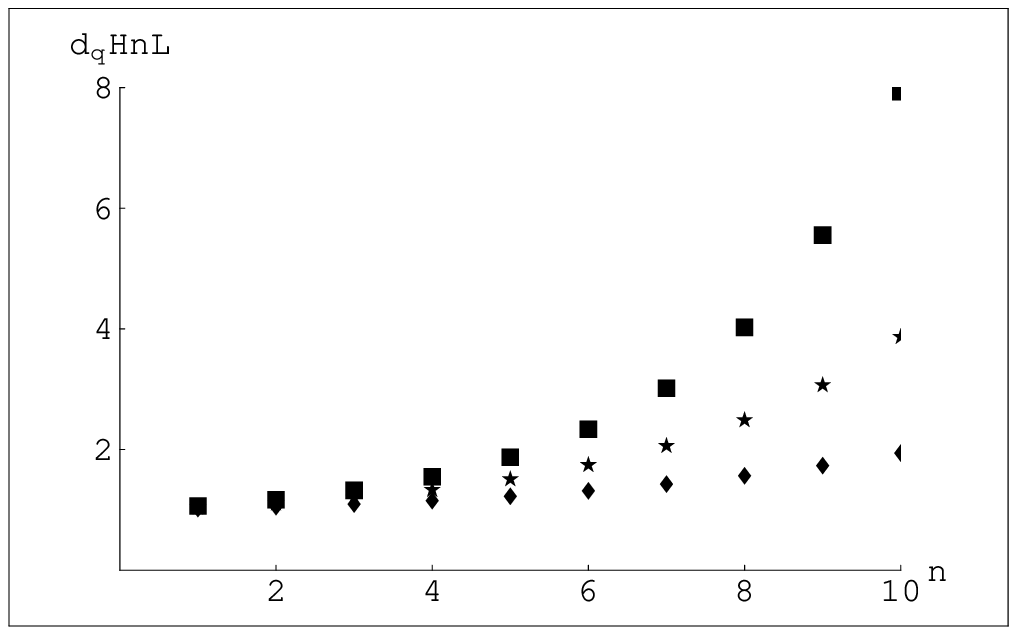}}}}
\end{picture}
\vspace{5cm}
\centerline{Figure 1}
%
%
\newpage
\begin{picture}(160,100)
\put(35,0){\mbox{\scalebox{1.0}{\includegraphics{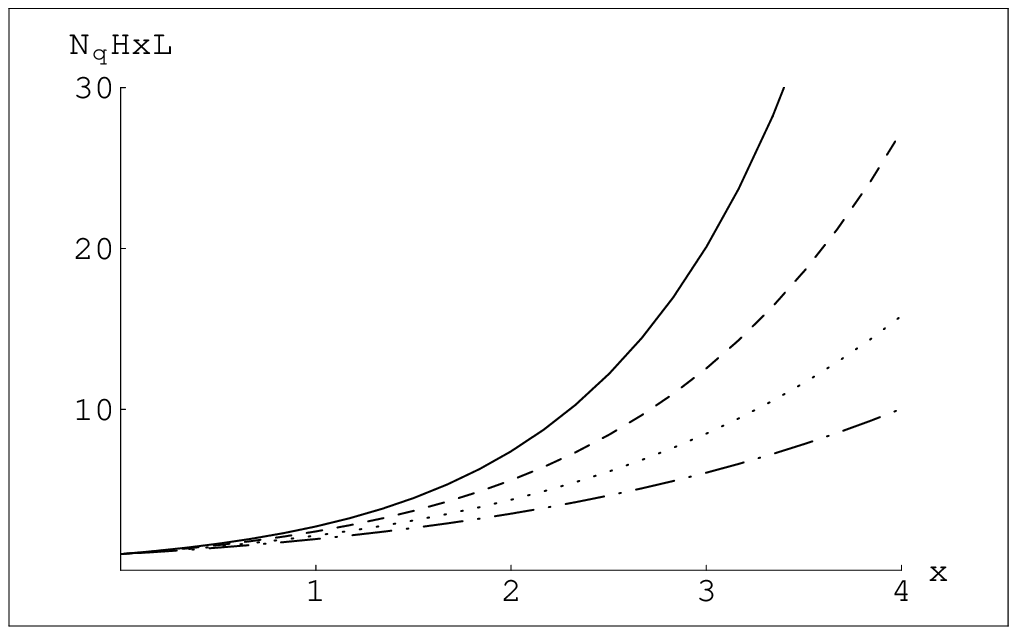}}}}
\end{picture}
\vspace{5cm}
\centerline{Figure 2}
%
%
\newpage
\begin{picture}(160,100)
\put(35,0){\mbox{\scalebox{1.0}{\includegraphics{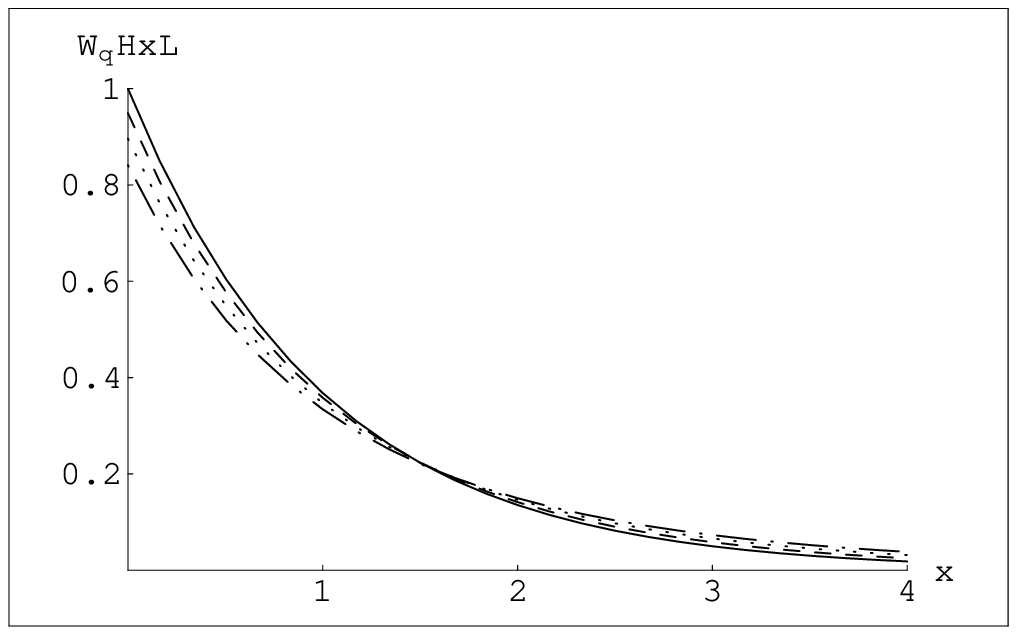}}}}
\end{picture}
\vspace{5cm}
\centerline{Figure 3}
%
%
\newpage
\begin{picture}(160,100)
\put(35,0){\mbox{\scalebox{1.0}{\includegraphics{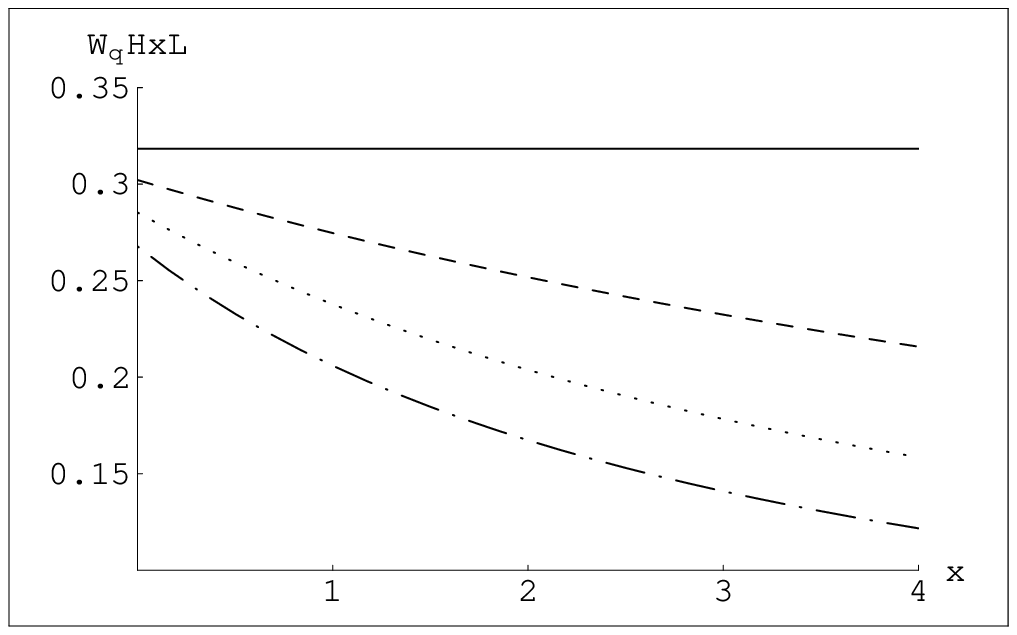}}}}
\end{picture}
\vspace{5cm}
\centerline{Figure 4}
%
%
\newpage
\begin{picture}(160,100)
\put(35,0){\mbox{\scalebox{1.0}{\includegraphics{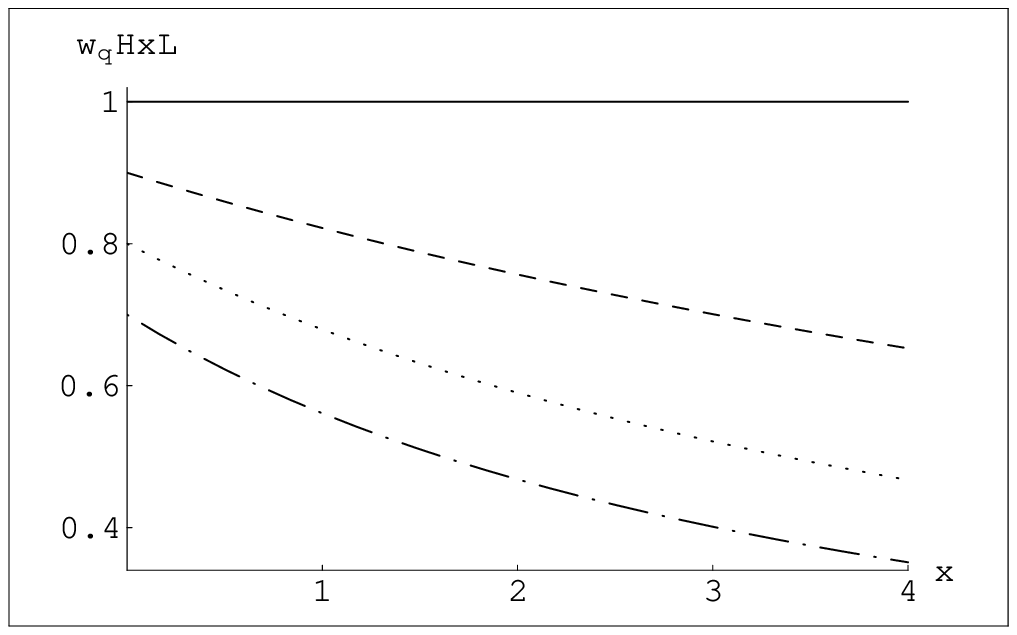}}}}
\end{picture}
\vspace{5cm}
\centerline{Figure 5}
%
%
\newpage
\begin{picture}(160,100)
\put(35,0){\mbox{\scalebox{1.0}{\includegraphics{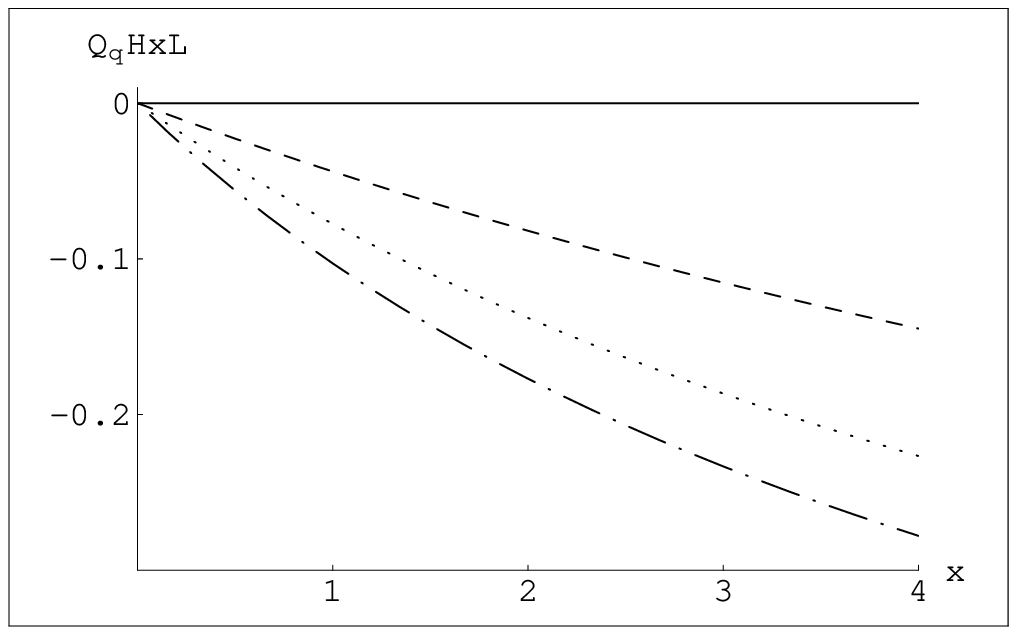}}}}
\end{picture}
\vspace{5cm}
\centerline{Figure 6}
%
%
\newpage
\begin{picture}(160,100)
\put(35,0){\mbox{\scalebox{1.0}{\includegraphics{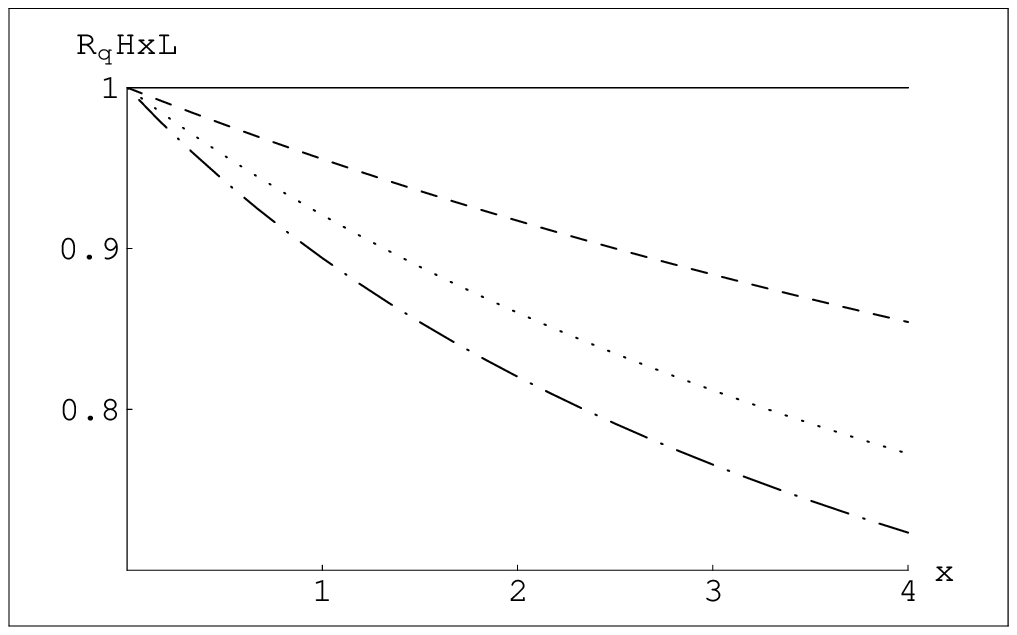}}}}
\end{picture}
\vspace{5cm}
\centerline{Figure 7}
%
%
\newpage
\begin{picture}(160,100)
\put(35,0){\mbox{\scalebox{1.0}{\includegraphics{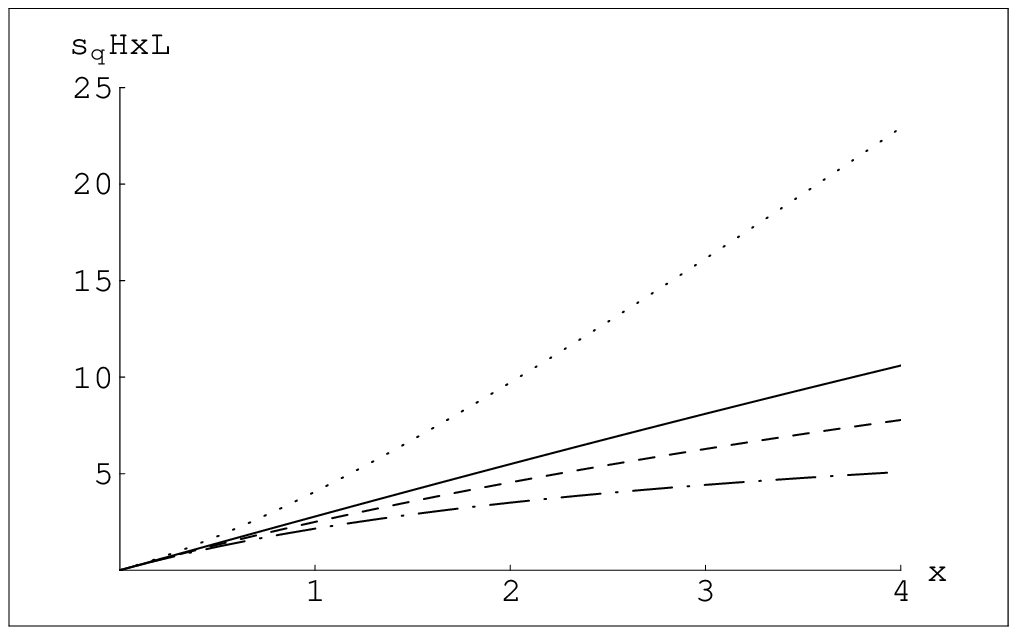}}}}
\end{picture}
\vspace{5cm}
\centerline{Figure 8}
%
%
\newpage
\begin{picture}(160,100)
\put(35,0){\mbox{\scalebox{1.0}{\includegraphics{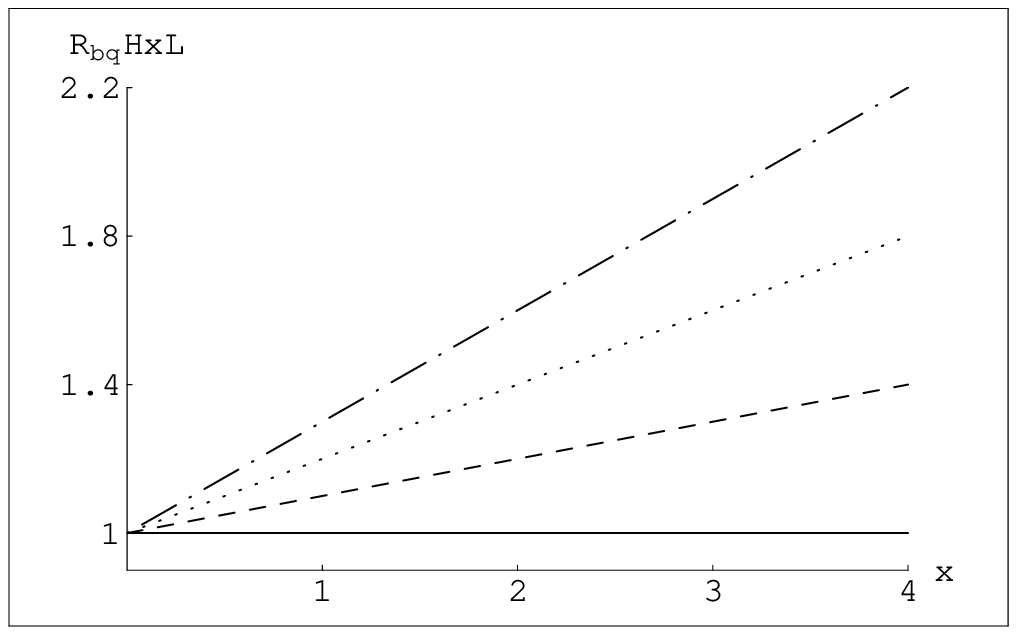}}}}
\end{picture}
\vspace{5cm}
\centerline{Figure 9}

\end{document}